\documentclass[english,prd,floatfix,superscriptaddress,nofootinbib]{revtex4-2}
\usepackage{graphicx} 
\usepackage{amsmath}
\usepackage[]{amssymb}
\usepackage[]{mathrsfs}
\usepackage[]{eucal}
\usepackage[]{tensor}
\usepackage[]{amsthm}
\usepackage[]{bm}
\usepackage{color}
\usepackage{array}
\usepackage{multirow}
\usepackage[margin=1in]{geometry}
\usepackage[normalem]{ulem}
\newcommand{\Epsilon}{\mathcal{E}}
\newcommand{\pf}[1]{\mathbf{#1}}

\newcommand{\hdg}{\star} 
\newcommand{\df}{\mathrm{d}}

\newcommand{\veps}{\bm{\epsilon}}

\newcommand{\bc}{\begin{center}}
\newcommand{\ec}{\end{center}}
\newcommand{\be}{\begin{equation}}
\newcommand{\ee}{\end{equation}}

\newcommand{\FF}{\mathcal{F}}
\newcommand{\GG}{\mathcal{G}}
\newcommand{\LL}{\mathscr{L}}
\usepackage[title]{appendix}

\begin{document}

\title{\bf Generalised Harrison transformations and black diholes in Einstein-ModMax}
\author{Ana Bokuli\'c}
\email{ana.bokulic@ua.pt}
\affiliation{Departamento de Matem\'atica da Universidade de Aveiro and Centre for Research and Development  in Mathematics and Applications (CIDMA), Campus de Santiago, 3810-193 Aveiro, Portugal}
\author{Carlos A. R. Herdeiro}
\email{herdeiro@ua.pt}
\affiliation{Departamento de Matem\'atica da Universidade de Aveiro and Centre for Research and Development  in Mathematics and Applications (CIDMA), Campus de Santiago, 3810-193 Aveiro, Portugal}
\affiliation{%
	Programa de Pós-Graduação em Física, Universidade Federal do Pará, 66075-110, Belém, Par{\'a}, Brazil 
}%

\begin{abstract}
Einstein-Maxwell theory has powerful solution generating techniques which include Harrison transformations in the Ernst formalism. We construct generalized Harrison transformations that preserve the purely magnetic or purely electric sector in Einstein-ModMax (EMM) theory. Thus, they serve as solution generating techniques within these sectors for this model of non-linear electrodynamics minimally coupled to gravity. As an application we rederive several known exact solutions of EMM and a new solution, black diholes, describing two extremal BHs in equilibrium, with opposite magnetic charges, whose attraction is balanced by their embedding in the Melvin magnetic universe of this model. As a further generalization, we consider Einstein-dilaton-ModMax theory, and provide the extremal charged BHs and black diholes also in this model.
\end{abstract}

\date{July 2025}

\maketitle

\section{Introduction}
Due to the inherently nonlinear character of Einstein's gravitational field equations, finding exact solutions in closed form poses a significant challenge. Therefore, solution-generating techniques that circumvent the need to solve the equations of motion directly are of considerable interest. A canonical example is the Ernst formalism \cite{PhysRev.168.1415} applied to the Einstein-Maxwell theory. Under the assumptions of stationarity and axial symmetry, the corresponding equations of motion can be reformulated in terms of complex gravitational and electromagnetic Ernst potentials. This framework reveals the symmetry structure of the theory that is otherwise not manifest, yet essential for generating new solutions from known ones. These symmetries form the Kinnersley group \cite{10.1063/1.1666373, 10.1063/1.523458}, which consists of two main classes of transformations: gauge transformations, which leave the seed spacetime physically unchanged, and two transformations that yield distinct solutions. The latter, known as the Ehlers \cite{Ehlers:1957zz} and Harrison \cite{10.1063/1.1664508} transformations, have been successfully utilised to construct a rich spectrum of exact solutions in Einstein-Maxwell theory. Depending on the type of metric ansatz used for a stationary and axially symmetric spacetime, their action on the seed solution can be summarised as follows: Ehlers transformation either adds a NUT charge or embeds the initial solution in a swirling background, while Harrison transformation immerses the spacetime into an electric or magnetic Melvin universe \cite{MELVIN196465} or introduces electromagnetic charges. One of the earliest applications of these methods was generating rotating NUT solutions \cite{10.1063/1.522614}, followed by embedding static and rotating (electro)vacuum solutions in magnetic Melvin universe \cite{1976JMP17182E, Ernst1976mzr}. 

What motivated further exploration of the transformations is the possibility of removing conical singularities in seed solutions. For example, the interaction between the external Melvin field and the charges of the C-metric solution introduces acceleration, effectively replacing struts needed to sustain the configuration \cite{PhysRevD.89.044022, 10.1063/1.522935, PhysRevD.82.024006}. A similar balancing mechanism can arise from spin-spin interactions between rotating BHs in swirling universes \cite{Astorino2022, Astorino:2025zse}. Harrison transformation has been successfully utilised to balance the Bonnor solution \cite{Bonnor1966}, allowing its reinterpretation as a pair of extremal, oppositely charged magnetic BHs in a magnetic universe, referred to as a \textit{dihole} \cite{EMP00}. The equilibrium is achieved by properly tuning the value of the external magnetic field so that the attraction between the black holes is counterbalanced. The generalisation to Einstein-Maxwell-dilaton theory \cite{EMP00} follows from the modified Harrison transformation \cite{PhysRevD.49.2909}, while the rotating dihole counterparts were presented in \cite{PhysRevD.90.024013}. Recently, starting from seed Kerr metric, a novel rotating solution to Einstein's vacuum equations has been obtained by performing a discrete "inversion" transformation of the gravitational Ernst potential  \cite{barrientos2025newexactrotatingspacetime}. A particularly intriguing property of this solution is the absence of both curvature and conical singularities.

Since the Ernst formalism has proven to be a fruitful method of obtaining solutions in Einstein-Maxwell theory, it is worth exploring the possibility of extending it to nonlinear generalisations of Maxwell’s electrodynamics. The term nonlinear electrodynamics (NLE) usually refers to Lagrangian densities which are functions of two electromagnetic invariants, $\tensor{F}{_a_b}\tensor{F}{^a^b}$ and $\tensor{F}{_a_b}{\hdg\tensor{F}{^a^b}}$. Generally, the structure of the equations of motion stemming from such a Lagrangian will be involved and therefore not easily expressed via Ernst potentials. However, a promising candidate is ModMax electrodynamics \cite{BLST20}, which bears similarity with Maxwell’s theory due to its conformal and electromagnetic duality symmetry. These properties make it especially interesting as a testing ground for studying the impact of NLE fields on exact black hole solutions, which have been expanding in recent years. A parallel between the two electromagnetic theories is evident in static solutions; spherically symmetric static ModMax black holes \cite{FAGMLM20} maintain a Maxwell-like form. A similar observation can be made for accelerated C-metric ModMax black holes \cite{BARRIENTOS2022137447} and ModMax-multi-black hole spacetimes \cite{PhysRevD.111.064046}. However, a significant deviation from Maxwell theory can be noticed in the gauge sector of the nonstatic Taub-NUT solution \cite{BORDO2021136312, Flores-Alonso2021}, as well as in nonstatic configurations with a conformally coupled scalar field \cite{PhysRevD.110.064027}, while the challenge of obtaining a rotating black hole still persists. The solutions presented in \cite{BCPK24}, including the ModMax Melvin universe and Schwarzschild and C-metric BHs embedded in it, suggest the existence of an underlying structure that emerges from a Harrison-like transformation. Guided by these insights, we will study the EMM equations and their corresponding symmetries in Ernst form. As it turns out, there are generalized Harrison transformations that preserve the purely magnetic or purely electric sector in Einstein-ModMax (EMM) theory and that we shall exploit to generate both known (obtained by other methods) and new solutions.

The paper is organised as follows. In Section \ref{sec2}, we provide a brief overview of the fundamentals of NLE theories and introduce ModMax electrodynamics. In the next section, we show how the Ernst scheme can be applied to the EMM framework and analyse the limitations of the formalism. In Section \ref{secMH}, we present the two distinct types of generalised Harrison transformations. In Section \ref{sec5}, we use the so-called magnetic Harrison transformation to derive a novel, dihole solution in ModMax theory and analyse its properties. Finally, Section \ref{sec6} contains a discussion and explores possible future applications of the Ernst-ModMax formalism.

\section{ModMax theory in a nutshell}\label{sec2}
Let us begin by introducing the basic objects of general NLE theories minimally coupled to gravity. Consider the Lagrangian consisting of the Einstein-Hilbert term and an NLE contribution,
\begin{align}\label{eq:Lagr}
\textbf{L}=\dfrac{1}{16\pi}(R+4\LL(\FF, \GG))\veps\ ,
\end{align}
where the electromagnetic invariants are constructed from the 2-form ${\textbf{F}=\df\textbf{A}}$ and its Hodge dual,
\begin{align}
\FF=F_{ab}F^{ab}\ \ \text{and}\ \ \GG=F_{ab}\,{\hdg F^{ab}}\ .
\end{align}
The equations of motion emerging via variational principle from (\ref{eq:Lagr}) are
\begin{equation}
\left.\begin{aligned}
R_{ab}-\dfrac{1}{2}&Rg_{ab}=8\pi T_{ab}\ ,\\
\df{\hdg\textbf{Z}}=0\ & \ \text{and}\ \ \df\textbf{F}=0\ ,
\end{aligned}\right.
\end{equation}
where $\textbf{Z}=-4(\partial_\FF\LL\textbf{F}+\partial_\GG\LL{\hdg\textbf{F}})$, while the energy-momentum tensor in a four-dimensional spacetime can be split into a "Maxwell-like" and a trace term,
\begin{align}\label{eq:emtensor}
T_{ab}=-4\partial_\FF\LL T_{ab}^{(Max)}+\dfrac{1}{4}T g_{ab}\ .
\end{align}
Along with the electric field and magnetic induction (4-vectors) defined with respect to a vector field $X^a$ as 
\begin{align}
\textbf{E}=-i_X\textbf{F}\ \  \text{and}\ \  \textbf{B}=i_X{\hdg\textbf{F}}\ ,
\end{align} 
respectively, one may introduce electric induction and magnetic field using the 2-form $\textbf{Z}$, 
\begin{align}\label{eq:DH}
\textbf{D}=-i_X\textbf{Z}\ \  \text{and}\ \  \textbf{H}=i_X{\hdg\textbf{Z}}\ .
\end{align}  
Komar electric and magnetic charges in the presence of nonlinear fields become
\begin{align}\label{eq:Komar}
Q=\dfrac{1}{4\pi}\oint_\mathcal{S} {\hdg\pf{Z}}\ \  \text{and}\ \  P=\dfrac{1}{4\pi}\oint_\mathcal{S}\pf{F}.
\end{align}
Next, we apply the above setup to nonlinear ModMax electrodynamics, defined by the following Lagrangian \cite{BLST20},
\begin{align}
\mathcal{L}^{(MM)}=\dfrac{1}{4}\left(-\FF\text{cosh}\gamma+\sqrt{\FF^2+\GG^2}\text{sinh}\gamma\right) ,
\end{align}
where the parameter $\gamma$ must be nonnegative for the theory to be causal. Notice that the ModMax Lagrangian does not respect Maxwell's weak field limit in a sense that $\LL\to-\FF/4$ as $\FF,\GG\to0$ does not hold. Maxwell's electrodynamics is instead recovered in the limit $\gamma\to0$. The parameter $\gamma$ effectively acts as a charge screening factor in the Coulomb solution or static black hole solutions \cite{FAGMLM20}. The explicit evaluation of the 2-form $\textbf{Z}$ yields
\begin{align}
\textbf{Z}=\left(\text{cosh}\gamma-\dfrac{\FF\,\text{sinh}\gamma}{\sqrt{\FF^2+\GG^2}}\right)\textbf{F}-\dfrac{\GG\,\text{sinh}\gamma}{\sqrt{\FF^2+\GG^2}}{\hdg\textbf{F}}\ .
\end{align}
Conformal invariance of the theory manifests as the vanishing trace of the energy-momentum tensor. Therefore, it can be immediately seen from (\ref{eq:emtensor}) that the energy-momentum tensor attains a simplified form,
\begin{align}
T_{ab}=\dfrac{1}{4\pi}\left(\text{cosh}\gamma-\dfrac{\FF}{\sqrt{\FF^2+\GG^2}}\text{sinh}\gamma\right) \left(\tensor{F}{_a_c}\tensor{F}{_b^c}-\dfrac{1}{4}\tensor{g}{_a_b}\FF\right)\ .
\end{align}
Another important property shared by ModMax theory and Maxwell's electrodynamics is the SO(2) duality invariance \cite{GR95}. The equations of motion and the energy-momentum tensor remain invariant under the rotations given by
\begin{align}
\widetilde{\textbf{F}}=\text{sin}\alpha\,{\hdg \textbf{Z}}+\text{cos}\alpha\, \textbf{F}\ ,\label{eq:so(2)1}\\
\widetilde{\textbf{Z}}=\text{cos}\alpha \,\textbf{Z}+\text{sin}\alpha\,{\hdg \textbf{F}}\ \label{eq:so(2)2}.
\end{align}
This symmetry can be exploited to generate dyonic solutions from the known electric or magnetic ones \cite{PhysRevD.111.064046}.
\section{Einstein-ModMax equations in Ernst formalism}\label{sec3}
Unlike in the case of Maxwell electrodynamics,  the Ernst scheme cannot be fully generalised to ModMax theory. However, there are two scenarios which allow description in terms of Ernst potentials. The key observation is the following: as long as the system has a nonvanishing electric or magnetic field only,  i.e. $\GG=0$, the ModMax theory is Maxwell-like.  Then, the equations of motion reduce to
\begin{equation}
\left.\begin{aligned}
&\tensor{R}{_a_b}=8\pi e^{\pm\gamma}T_{ab}^{(Max)}\ ,\\
&\df{\hdg\textbf{F}}=0\ \ \text{and}\ \ \df\textbf{F}=0\ ,
\end{aligned}\right.
\end{equation}
where $\pm$ sign corresponds to electrically and magnetically charged configurations, respectively. In principle, the $\GG=0$ condition could be relaxed by requiring solely proportionality between the invariants $\FF$ and $\GG$, as the linearity of the theory would still be preserved \cite{BCPK24}. We will provide a discussion of this generalisation in Appendix \ref{app}. Furthermore, we restrict our attention to static spacetimes, as the presence of rotation typically generates both electric and magnetic fields (e.g. NUT BHs), which, in a general case, could violate the proportionality condition $\FF\sim\GG$. The metric of a static and axisymmetric spacetime may be put either in the "electric" Lewis-Weyl-Papapetrou (LWP) form, which in cylidrical coordinates reads
\begin{align}\label{eq:LWPE}
ds^2=\dfrac{1}{f(\rho, z)}[\rho^2\df \phi^2+e^{2k(\rho, z)}(\df\rho^2+\df z^2)]-f(\rho, z)\df t^2 \ ,
\end{align}
or alternatively, in the "magnetic" LWP form
\begin{align}\label{eq:LWPB}
ds^2=\dfrac{1}{f(\rho, z)}[-\rho^2\df t^2+e^{2k(\rho, z)}(\df\rho^2+\df z^2)]+f(\rho, z)\df \phi^2 \ .
\end{align}
The ansatz for the gauge field that accommodates either a purely electric 
\begin{align}
\textbf{A}=A_t\df t\ ,
\label{electric_A}
\end{align}
or a purely magnetic field 
\begin{align}
\textbf{A}=A_\phi\df\phi\ .
\label{magnetic_A}
\end{align}
We remark that the electric form of the metric ansatz~\eqref{eq:LWPE}, can be combined with both a purely electric~\eqref{electric_A} or a purely magnetic~\eqref{magnetic_A} gauge ansatz. The same applies to the magnetic form of the metric ansatz~\eqref{eq:LWPB}.   
Thus, the terms 'electric' and 'magnetic' in the metric ansatz should not be associated to the nature of the gauge field; they are useful to distinguish the action of the symmetry transformations on seed solutions, which will be discussed in Section \ref{secMH}. 
Observe that the transition from the electric to the magnetic form is obtained by a Wick-like rotation, $t\to i\phi$ and $\phi\to it$, accompanied by the same transformation of the electromagnetic potential.  

To translate the EMM field equations into the Ernst framework, we express the metric function $f$ and the gauge field $\textbf{A}$ in terms of the complex potentials $\Epsilon$ and $\Phi$. The exact relations depend both on the type of metric ansatz (electric or magnetic) and on the type of gauge field ansatz (electric or magnetic). For the
\begin{description}
\item[i)] electric metric and electric gauge field ansatz:
\begin{align}\label{eq:EE}
f=\Epsilon+\Phi^{*}\Phi e^{\gamma}, \hspace{5mm} \Phi=A_t\ \ ;
\end{align}
\item[ii)] electric metric and magnetic gauge field ansatz:
\begin{align}\label{eq:EB}
f=\Epsilon+\Phi^{*}\Phi e^{-\gamma}, \hspace{5mm} \Phi=i\tilde{A_\phi}\ ,
\end{align}
where $\tilde{A}_\phi$ is the auxiliary electromagnetic twisted potential, from which the actual gauge potential ($A_\phi$) may be extracted via
\begin{align}\label{eq:EBA}
\hat{\phi}\times\nabla\tilde{A}_\phi=\dfrac{f}{\rho}\nabla A_\phi\ ;
\end{align}
where $\hat{\phi}$ is the azimuthal unit vector. 
\item[iii)] magnetic metric and magnetic gauge field ansatz:
\begin{align}\label{eq:BE}
f=-\Epsilon-e^{-\gamma}\Phi^*\Phi\ , \hspace{5mm}\Phi=-iA_\phi \ ;
\end{align}
\item[iv)] magnetic metric and electric gauge field ansatz:
\begin{align}
f=-\Epsilon-e^{\gamma}\Phi^*\Phi\ , \hspace{5mm}\Phi=\tilde{A_t} \ ,\label{eq:BB}\\
\hat{\phi}\times\nabla\tilde{A}_t=-\dfrac{f}{\rho}\nabla A_t\label{eq:BBA}\ .
\end{align}
\end{description}

We remark that the Ernst potential $\Epsilon$ is real in the static case, since its imaginary part is related to rotation. The metric function $k(\rho, z)$ is defined by a system of partial differential equations in $f(\rho, z)$ and gauge potential, given by rather lengthy expressions in Ernst formalism. We do not present them here as our focus will be on the Ernst transformations of existing solutions, which leave the function $k(\rho, z)$ unaltered. 

In the language of the Ernst formalism, the EMM equations for both the electric and magnetic metric ansatz but with a purely electric gauge field become
\begin{equation}
\left.\begin{aligned}\label{eq:ErnstE}
(\Epsilon+e^{\gamma}\Phi^*\Phi)\nabla^2\Epsilon={\nabla\Epsilon\cdot}(\nabla\Epsilon+2e^{\gamma}\Phi^*\nabla\Phi)\ ,  \\
(\Epsilon+e^{\gamma}\Phi^*\Phi)\nabla^2\Phi={\nabla\Phi\cdot}(\nabla\Epsilon+2e^{\gamma}\Phi^*\nabla\Phi)\ ,
\end{aligned}\right.
\end{equation}
while for both the electric and magnetic metric ansatz but with a purely magnetic gauge field read
\begin{equation}
\left.\begin{aligned}\label{eq:ErnstB}
(\Epsilon+e^{-\gamma}\Phi^*\Phi)\nabla^2\Epsilon={\nabla\Epsilon\cdot}(\nabla\Epsilon+2e^{-\gamma}\Phi^*\nabla\Phi)\ ,  \\
(\Epsilon+e^{-\gamma}\Phi^*\Phi)\nabla^2\Phi={\nabla\Phi\cdot}(\nabla\Epsilon+2e^{-\gamma}\Phi^*\nabla\Phi)\ .
\end{aligned}\right.
\end{equation}

In the next section, we show how the presented scheme enables generating new purely electric or purely magnetic solutions (in what concerns the gauge field) in EMM theory by relying on the symmetries of modified Ernst equations. 
\section{Modified Harrison transformation}\label{secMH}
As emphasised earlier, Ernst formalism reveals a powerful solution-generating technique for Einstein-Maxwell theory. In ModMax theory, however, the symmetry is reduced and is only realised in special cases. First, it's important to distinguish pure gauge transformations from nontrivial ones. For example, it can be shown that the transformations of the form
\begin{align}
\Epsilon'=\Epsilon-2e^{\pm\gamma}\lambda^*\Phi-|\lambda|^2e^{\pm\gamma}\, , &\ \ \Phi'=\Phi+\lambda\label{eq:EMgauge}\ ,
\end{align}
where $\lambda\in\mathbb{C}$, $\lambda^*\Phi\in\mathbb{R}$ and $\pm$ sign denotes electric and magnetic cases, represent the electromagnetic gauge transformations and thus do not produce new independent solutions.

To draw a parallel between Maxwell and ModMax theories, we had to impose a rather restrictive staticity condition that must be preserved even after applying tranformations. Its further consequence is that it automatically makes Ehlers transformations, which add the imaginary component to the transformed potential $\Epsilon'$ \cite{PhysRevD.108.024059}, generally inapplicable in the ModMax case. The known nonstatic Taub-NUT \cite{Flores-Alonso2021,BORDO2021136312} and swirling universe solutions \cite{BCPK24} in ModMax theory support this claim. These solutions cannot emerge from an Ehlers-like transformation as the electromagnetic invariants are not proportional, causing the system of equations to deviate significantly from the Maxwell case.  The same proportionality condition is violated when an electrically charged seed solution is placed in the magnetic Melvin universe \cite{Gibbons_2013, PhysRevD.89.044022} \footnote{or vice versa, if the magnetic seed is embedded in the electric universe}, where the interaction of the external magnetic field and the initial electric field gives rise to the twist potential. However, the appropriately modified Harrison transformation can still be used as a solution-generating technique, provided that the gauge potential is purely real (electric) or purely imaginary (magnetic).  This leaves us with several options for a solution generating technique in EMM, based on generalised Harrison transformations, defined with parameter $\alpha$. One can take the seed solution to be: 
\begin{description}
    \item[A)] a vacuum metric in electric LWP form and keeping $\alpha$ either real or imaginary;
    \item[BI)] a vacuum metric in magnetic LWP form and keeping $\alpha$ either real or imaginary;
    \item[BII)] a non-vacuum metric in magnetic LWP form, with an electric gauge ansatz and keeping $\alpha$ real;
      \item[BIII)] a non-vacuum metric in magnetic LWP form, with a magnetic gauge ansatz and keeping $\alpha$ imaginary;
\end{description} 

The generalised Harrison transformation that leaves the system of equations (\ref{eq:ErnstE}) invariant, thus moving along orbits of the purely electric sector (in what concerns the gauge field) of EMM solutions, is given by
\begin{align}\label{eq:HarrE}
\Epsilon'=\dfrac{\Epsilon}{1-2\alpha^*\Phi -\alpha^*\alpha\Epsilon e^{-\gamma}}\ , \ \ \Phi'=\dfrac{\alpha\Epsilon e^{-\gamma}+\Phi}{1-2\alpha^*\Phi -\alpha^*\alpha\Epsilon e^{-\gamma}}\ ,
\end{align}
while the one that preserves the system (\ref{eq:ErnstB}), thus moving along orbits of the purely magnetic sector (in what concerns the gauge field) of EMM solutions, is 
\begin{align}\label{eq:HarrB}
\Epsilon'=\dfrac{\Epsilon}{1-2\alpha^*\Phi e^{-\gamma}-\alpha^*\alpha\Epsilon e^{-\gamma}}\ , \ \ \Phi'=\dfrac{\alpha\Epsilon+\Phi}{1-2\alpha^*\Phi e^{-\gamma}-\alpha^*\alpha\Epsilon e^{-\gamma}}\ .
\end{align}
We will refer to the transformations (\ref{eq:HarrE}) and (\ref{eq:HarrB}) as electric and magnetic Harrison transformations, respectively. In the following sections, we discuss the action of the modified Harrison transformation on different seed solutions and provide examples for each scenario.

\subsection{Electric LWP ansatz}
The known solutions for static, electrically charged EMM BHs were derived by solving the equations of motion directly using the appropriate ansatz \cite{FAGMLM20}.  The Ernst scheme provides an alternative method of obtaining them and other possible electrically charged, static configurations. The basic strategy is the following: start from a vacuum seed solution with $A_t=A_\phi=0$ written in the electric LWP form so that $\Epsilon=f$ and apply either the electric or the magnetic Harrison transformation. 
To illustrate the idea, we start from the Schwarzschild solution adapted to the anstaz (\ref{eq:LWPE}). Taking into account the relation between sperical and cylindrical coordinates 
\begin{align}
\rho=\sqrt{r(r-2M)}\,\text{sin}\,\theta\ ,\ z=(r-M)\,\text{cos}\,\theta\ ,
\end{align}
we get
\begin{align}
\Epsilon=1-\dfrac{2M}{r}\ ,\ \ e^{2k}=\dfrac{r^2-2Mr}{r^2-2Mr+M^2\text{sin}^2\theta}\ .
\end{align}
After appliying the electric Harrison transformation (\ref{eq:HarrE}), 
\begin{align}
\Epsilon'=\dfrac{r-2M}{r-\alpha^2e^{-\gamma}(r-2M)}\ , \ \Phi'=\dfrac{e^{-\gamma}\alpha(r-2M)}{r-\alpha^2e^{-\gamma}(r-2M)}   \ ,
\end{align}
where the parameter $\alpha$ is taken to be real, the new metric function and gauge potential calculated from (\ref{eq:EE}) are given by
\begin{align}
f'=\dfrac{r(r-2M)}{(r-\alpha^2e^{-\gamma}(r-2M))^2}\ , \hspace{5mm} A'_t=\dfrac{e^{-\gamma}\alpha(r-2M)}{r-\alpha^2e^{-\gamma}(r-2M)}\ .
\end{align}
The following coordinate changes and parameter redefinitions,
\begin{equation}
\left.\begin{aligned}\label{eq:coord1}
r=\dfrac{\bar{r}-2M\alpha^2 e^{-\gamma}}{1-\alpha^2 e^{-\gamma}}\ , \ &Q=2\alpha M, \ \ m=M(1+\alpha^2 e^{-\gamma})\ ,\\
dr=\dfrac{d\bar{r}}{1-\alpha^2 e^{-\gamma}},&\ \ t=\bar{t}(1-\alpha^2 e^{-\gamma})\ ,
\end{aligned}\right.
\end{equation}
allow one to recoginse the electrically charged ModMax black hole in standard spherical coordinates,
\begin{align}\label{eq:MME}
ds^2=-\left(1-\dfrac{2m}{\bar{r}}+\dfrac{Q^2e^{-\gamma}}{\bar{r}^2} \right)\df \bar{t}^2+\left(1-\dfrac{2m}{\bar{r}}+\dfrac{Q^2e^{-\gamma}}{\bar{r}^2} \right)^{-1}\df \bar{r}^2+ \bar{r}^2\df \theta^2+\bar{r}^2\text{sin}^2\theta\df \phi^2\ ,\ A'_{\bar{t}}=-\dfrac{Qe^{-\gamma}}{\bar{r}}\ .
\end{align}
Similarly, to add the magnetic charge to Schwarzschild seed solution, one uses the magnetic Harrison transformation (\ref{eq:HarrB}) with imaginary $\alpha$, resulting in
\begin{align}
f'=\dfrac{r(r-2M)}{(r-|\alpha|^2e^{-\gamma}(r-2M))^2}\ , \hspace{5mm} \tilde{A'}_{\phi}=\dfrac{\text{Im}\alpha(r-2M)}{r-|\alpha|^2 e^{-\gamma}(r-2M)}\ .
\end{align}
The gauge potential can be calculated from (\ref{eq:EBA}) using $\nabla\propto\hat{r}\sqrt{r^2-2Mr}\partial_r+\hat{\theta}\partial_\theta$,
\begin{align}
A'_{\phi}=-2\text{Im}\alpha M\text{cos}\theta\ .
\end{align}
The same coordinate changes (\ref{eq:coord1}) with $P=2\text{Im}\alpha M$ reproduce the magnetically charged EMM BH,
\begin{align}
ds^2=-\left(1-\dfrac{2m}{\bar{r}}+\dfrac{P^2e^{-\gamma}}{\bar{r}^2} \right)\df t^2+\left(1-\dfrac{2m}{\bar{r}}+\dfrac{P^2e^{-\gamma}}{\bar{r}^2} \right)^{-1}\df \bar{r}^2+ \bar{r}^2\df \theta^2+\bar{r}^2\text{sin}^2\theta\df \phi^2\ ,\ A'_\phi=-P\cos\theta\ .
\end{align}
Due to the duality invariance of the theory, the dyonic solution is obtained by superposing the gauge potentials, $\textbf{A}=A_t \df t+A_\phi\df \phi$, and redefining the charge in metric functions to $Q^2+P^2$. Duality invariance also enables straightforward generalisation from electric to dyonic solution \cite{PhysRevD.111.064046}, therefore circumventing the use of magnetic Harrison transformation.  

Other solutions that could be generated using this approach include accelerated, non-rotating EMM BHs. After applying the Harrison transformation to the C-metric solution, the resulting spacetime is expected to belong to the Petrov type-I class \cite{PhysRevD.108.124025, Barrientos2024}, and therefore, differ from the C-metric accelerated ModMax BHs presented in \cite{BARRIENTOS2022137447}.

\subsection{Magnetic LWP ansatz}
{\bf I)} Consider the action of Harrison transformations on the vacuum seed solutions. Upon setting $\alpha=-iB/2$, where $B$ is the external magnetic field parameter, the magnetic transformation (\ref{eq:HarrB}) of the metric components and the gauge field may be written compactly as 
\begin{align}
g'_{ii}=\lambda^2 g_{ii}\ \  \text{if}\ \  i\neq\phi\ , \  g'_{\phi\phi}=\lambda^{-2}g_{\phi\phi}\  ,\ \ A'_\phi=\dfrac{-2e^{\gamma}(-1+\lambda)}{B\lambda}\ ,
\end{align}
with $\lambda=1+B^2e^{-\gamma}f(\rho, z)/4$.  Analogously, if we choose $\alpha=D/2$, where $D$ parametrises the strength of the electric induction defined in (\ref{eq:DH}),  the electric transformation (\ref{eq:HarrE}) takes the form
\begin{align}
g'_{ii}=\lambda^2 g_{ii}\ \  \text{if}\ \  i\neq\phi\ , \  g'_{\phi\phi}=\lambda^{-2}g_{\phi\phi}\  ,\ \ \tilde{A'}_t=\dfrac{-2(-1+\lambda)}{D\lambda}\ ,
\end{align}
where $\lambda=1+D^2e^{-\gamma}f(\rho, z)/4$. To check the consistency of this method, one may reproduce results from \cite{BCPK24} and derive the Melvin-ModMax universe, as well as Schwarzschild and C-metric solutions embedded in Melvin-ModMax theory.  

As a particular example, we may again take the Schwarzschild solution as a seed, albeit written in a magnetic LWP form so that $f=r^2\text{sin}^2\theta$.  The solution representing Schwarzschild BH in the Melvin universe is given by
\begin{equation}
\left.\begin{aligned}
\df s^2&=-\lambda^2 f\df t^2+\dfrac{\lambda^2}{f}\df r^2+r^2\lambda^2 \df\theta^2+r^2\text{sin}^2\theta\lambda^{-2}\df\phi^2, \ \ f=1-\dfrac{2M}{r}\ ,\\
\textbf{A}'&=-\dfrac{Br^2\text{sin}^2\theta}{2\lambda}\df\phi\ ,
\end{aligned}\right.
\end{equation}
where $\lambda=1+B^2e^{-\gamma}r^2\text{sin}^2\theta/4$. The solution reduces to Melvin-ModMax universe when $M=0$.

So far, we have considered magnetic universe only, which was achieved by keeping the parameter $\alpha$ imaginary. It is possible to introduce the electric component by relying on the duality invariance of the theory.  From the transformed electromagnetic form (\ref{eq:so(2)1}),
\begin{align}
\widetilde{\textbf{F}}=\text{cos}\,\alpha\textbf{F}+e^{-\gamma}\text{sin}\,\alpha{\hdg\textbf{F}}\ ,
\end{align}
we can extract the gauge potential
\begin{align}
\textbf{A}=-e^{-\gamma}\widetilde{D}rf\cos\theta\df t-\dfrac{\widetilde{B}r^2\text{sin}^2\theta}{2\lambda}\df\phi\ ,
\end{align}
where $\widetilde{B}=\cos\alpha B$ and $\widetilde{D}=-\sin\alpha B$. In the metric, the change amounts to replacing the square of magnetic field with $\widetilde{B}^2+\widetilde{D}^2$. Alternatively, the electric field can be generated by using the electric Harrison transformation (\ref{eq:HarrE}) and calculating the potential from (\ref{eq:BBA}), which yields
\begin{align}
A'_t=-De^{-\gamma}r\cos\theta f\ ,
\end{align}
in agreement with the previous result. However, from an astrophysical perspective, magnetic Melvin universe represents a more plausible model near BHs, as they are surrounded by strong magnetic fields emanating from accretion discs 
or from a wider galactic environment.

{\bf II)} Another possibility is to start from the electric seed $(A_t\neq0, A_\phi=0)$ and place it into the electric universe. In this case the transformation (\ref{eq:HarrE}) becomes
\begin{equation}
\left.\begin{aligned}
g'_{ii}=\lambda^2 g_{ii}\ \  \text{if}\ \ & i\neq\phi\ , \  g'_{\phi\phi}=\lambda^{-2}g_{\phi\phi},\ \tilde{A'}_t=\dfrac{-2(-1+\lambda)-D\tilde{A}_t}{D\lambda}\ ,\\
\lambda&=\left(1-\dfrac{1}{2}D\tilde{A}_t\right)^2+\dfrac{1}{4}D^2e^{-\gamma}f(\rho, z)\ .
\end{aligned}\right.
\end{equation}
As an example, we can derive a solution representing the electrically charged ModMax BH embedded in the electric universe.
From the initial metric (\ref{eq:MME}), we identify
\begin{align}
f=r^2\text{sin}^2\theta,\ \ \tilde{A}_t=-Qe^{-\gamma}\text{cos}\theta\ .
\end{align}
After applying the transformation, we have
\begin{equation}
\left.\begin{aligned}\label{eq:elel}
\df s^2&=-\lambda^2 f\df t^2+\dfrac{\lambda^2}{f}\df r^2+r^2\lambda^2 \df\theta^2+r^2\text{sin}^2\theta\lambda^{-2}\df\phi^2, \ \ f=1-\dfrac{2M}{r}+\dfrac{Q^2e^{-\gamma}}{r^2}\ ,\\
A'_t&=-\dfrac{Qe^{-\gamma}}{r}\left(1+\dfrac{1}{4}e^{-\gamma}D^2r^2+\dfrac{1}{4}D^2r^2e^{-\gamma}f\cos^2\theta\right)-e^{-\gamma}Dfr\cos\theta\ ,\\
\lambda&=\left(1+\dfrac{1}{2}DQe^{-\gamma}\text{cos}\theta\right)^2+\dfrac{1}{4}D^2e^{-\gamma}r^2\text{sin}^2\theta\ .
\end{aligned}\right.
\end{equation}
Unfortunately, this solution is plagued by conical singularities, which is quite natural as an electrically charged BH would accelerate in a background electric field. Along the axis,  for $\theta=0$ ($\theta=\pi$), there is a conical deficit (excess),
\begin{align}
\delta\big|_{\theta=0}=2\pi\left(1-\dfrac{16}{(2+e^{\gamma}DQ)^4}   \right)\, ,\ \delta\big|_{\theta=\pi}=2\pi\left(1-\dfrac{16}{(2-e^{-\gamma}DQ)^4}   \right)\ .
\end{align}
A more interesting example could be obtained by immersing the accelerated electrically charged ModMax BH into the electric universe. In Maxwell electrodynamics, this procedure removes conical singularities \cite{10.1063/1.522935, PhysRevD.82.024006}, indicating that the same might apply in the ModMax scenario. 

{\bf III)} Lastly, we start with the magnetic seed ($A_t=0$, $A_\phi\neq0$). Using simpler notation, the transformation (\ref{eq:HarrB}) that adds the magnetic Melvin background can be written as
\begin{equation}
\left.\begin{aligned}
g'_{ii}=\lambda^2 g_{ii}\ \  \text{if}\ \ & i\neq\phi\ , \  g'_{\phi\phi}=\lambda^{-2}g_{\phi\phi}\ ,\ \ 
A'_\phi=\dfrac{-2e^{\gamma}(-1+\lambda)-BA_\phi}{B\lambda}\ ,\\
\lambda&=\left(1-\dfrac{1}{2}e^{-\gamma}BA_\phi\right)^2+\dfrac{1}{4}B^2e^{-\gamma}f(\rho, z)\ .
\end{aligned}\right.
\end{equation}
In the next section,  we utilise this method to derive the ModMax dihole solution analogous to the one presented in \cite{EMP00}.

The following table summarizes the action of the Harrison transformations,  either electric (E) or magnetic (M), on seed BH spacetimes expressed in electric or magnetic ansatz discussed in this section.  The pairs of electrically and magnetically charged solutions in the final column, within two horizontal lines, are moreover related by electromagnetic duality symmetry. For completeness the last two lines make clear that starting with the electric/magnetic ModMax BH in the electric LWP metric ansatz and applying the electric/magnetic Harrison transformation is an identity map.

\setlength\extrarowheight{3pt}
\begin{center}
\begin{tabular}{ p{10em} | p{1.5cm} | p{2cm}  | p{3cm} | p{5.5cm} } 
\hline
\begin{minipage}[c][2.5em][c]{\linewidth}\centering Seed BH\end{minipage} &
\begin{minipage}[c][2.5em][c]{\linewidth}\centering Metric\\Ansatz\end{minipage} &
\begin{minipage}[c][2.5em][c]{\linewidth}\centering Gauge field\\ansatz\end{minipage} &
\begin{minipage}[c][2.5em][c]{\linewidth}\centering Harrison\\transformation\end{minipage} &
\begin{minipage}[c][2.5em][c]{\linewidth}\centering Result\end{minipage} \\
  \hline\hline
  Schwarzschild &\centering E &\centering None &\centering E &electric ModMax \\ 
  Schwarzschild & \centering E &\centering None & \centering M&magnetic ModMax \\ 
\hline
  Schwarzschild & \centering M &\centering None &\centering M&Schwarzschild-Melvin \\ 
  Schwarzschild &\centering M &\centering None &\centering E&Schwarzschild-electric universe\\
\hline
 electric ModMax &\centering M &\centering E &\centering E&electric ModMax-electric universe\\
 magnetic ModMax &\centering M &\centering M&\centering M&magnetic ModMax-Melvin\\
\hline
 electric ModMax &\centering E &\centering E &\centering E&electric ModMax\\
 magnetic ModMax &\centering E &\centering M&\centering M&magnetic ModMax\\
\hline
\end{tabular}\\ 
\vspace{2mm}
{\raggedright \par}
\end{center}

\section{Dihole solution in ModMax electrodynamics}\label{sec5}
We now turn our attention to the construction of a solution representing two extremal BHs, with opposite charge (thus a dipole or \textit{dihole}), in a Melvin universe. The final solution will be free of conical singularities. It is the EMM version of the Einstein-Maxwell solution in~\cite{EMP00}.

The construction of the solution is done in two steps; the first one consists of generalising the Bonnor dipole solution \cite{Bonnor1966} to ModMax theory and the second one makes use of the previously introduced Harrison transformation to embed it into an asymptotically Melvin universe.

Let us consider the first step. It has been shown that the Bonnor solution, which describes a mass with a magnetic dipole moment, can be in fact interpreted as two static, magnetically charged extremal BHs with opposite charges \cite{EMP00}.  The corresponding solution in ModMax theory is given by the following metric and gauge potential,
\begin{equation}
\left.\begin{aligned}\label{eq:BMM}
\df s^2&=-\left(1-\dfrac{2Mr}{\Sigma}\right)^2\df t^2+\dfrac{(1-\frac{2Mr}{\Sigma})^2\Sigma^4}{(\Delta+(M^2+a^2e^{-\gamma})\text{sin}^2\theta)^3}\left(\dfrac{\df r^2}{\Delta}+\df \theta^2\right)+
\dfrac{\Delta\text{sin}^2\theta}{(1-\frac{2Mr}{\Sigma})^2}\df \phi^2\ ,\\
\textbf{A}&=\dfrac{2aMr\text{sin}^2\theta}{\Delta+a^2e^{-\gamma}\text{sin}^2\theta}\df \phi\ ,
\end{aligned}\right.
\end{equation}
where we used the abbreviations $\Sigma=r^2-a^2\text{cos}^2\theta e^{-\gamma}$ and $\Delta=r^2-2Mr-a^2e^{-\gamma}$. Even though this solution was obtained through educated guesswork, there exists a systematic algorithm for generating static, axially symmetric configurations of the Einstein-Maxwell-(dilaton) equations, starting from stationary vacuum solutions \cite{DAVIDSON1994304, Liang:2001sp}. Using the notation of \cite{Liang:2001sp}, this procedure can be adapted to the ModMax case, provided that $w=i\sqrt{1+\alpha^2}e^{-\gamma/2}A_\phi$.

One feature of the solution (\ref{eq:BMM}) is a conical singularity along the segment of the symmetry axis defined by $r_+=M+\sqrt{M^2+a^2e^{-\gamma}}$,
\begin{align}
\delta=2\pi\left(1-\left(1+\dfrac{M^2e^{\gamma}}{a^2}\right)^2\right)\ .
\end{align} 
This localised strut source arises due to unbalanced gravitational attraction. One possible way to counterbalance it is by adding the repulsive interaction in the form of the external magnetic field, which renders the spacetime non-asymptotically flat, which, as described above, can be done by an appropriate Harrison transformation. 

Now we turn to the second step. The Bonnor-ModMax solution (\ref{eq:BMM}) serves as a seed solution, to which we apply the magnetic Harrison transformation (\ref{eq:HarrB}), thereby immersing it into the Melvin-ModMax universe. The resulting metric and gauge field are given by
\begin{equation}
\left.\begin{aligned}\label{eq:MMdihole}
\df s^2&=-\Lambda^2 \df t^2+\dfrac{\Lambda^2\Sigma^4}{(\Delta+(M^2+a^2e^{-\gamma})\text{sin}^2\theta)^3}\left(\dfrac{\df r^2}{\Delta}+\df \theta^2\right)+\dfrac{\Delta\text{sin}^2\theta}{\Lambda^2}\df\phi^2\ , \\
\textbf{A}&=-\dfrac{-2Mra+\frac{1}{2}B((r^2-a^2e^{-\gamma})^2+\Delta a^2e^{-\gamma}\text{sin}^2\theta)}{\Lambda\Sigma}\text{sin}^2\theta\df \phi\ ,
\end{aligned}\right.
\end{equation}
where
\begin{align}
\Lambda=\dfrac{\Delta+a^2e^{-\gamma}\text{sin}^2\theta-2BMrae^{-\gamma}\text{sin}^2\theta+\frac{1}{4}B^2e^{-\gamma}\text{sin}^2\theta((r^2-a^2e^{-\gamma})^2+\Delta a^2e^{-\gamma}\text{sin}^2\theta)}{\Sigma}\ .
\end{align}
The conical excess equal to
\begin{align}
\delta=2\pi\left(1-\left(1+\dfrac{M^2e^{\gamma}}{a^2}\right)^2\left(1-\dfrac{BMr_+}{a}\right)^{-4}    \right)\ ,
\end{align}
is cancelled if the value of the magnetic field is fune-tuned to 
\begin{align}\label{eq:mag}
B=\dfrac{a\pm\sqrt{a^2+e^{\gamma}M^2}}{Mr_+}\ .
\end{align}
In the limit of infinite separation between the two BHs, indicated by $a$, we expect the magnetic field to vanish. Therefore, we choose the minus sign since in that case $B\to-e^{\gamma/2}M/(2a^2e^{-\gamma})$ as $a\to\infty$.

To interpret that this solution corresponds to ModMax-dihole configuration, we will follow the approach from \cite{EMP00}. In the limit $r=r_+$ and $\theta\in\{0,\pi\}$, the metric becomes singular. However, when expressed in appropriate coordinates,  these points describe the-near horizon geometry of extremal ModMax BHs. To see this, we set $r=r_+ +\rho(1+\cos\bar{\theta})/2$ and $\sin^2\theta=\rho(1-\cos\bar{\theta})/\sqrt{M^2+a^2e^{-\gamma}}$. Around $r=r_+$ and $\theta=0$, the solution becomes
\begin{align}
ds^2&=-\dfrac{\rho^2}{Q^2}\df t^2+\dfrac{Q^2}{\rho^2}\df \rho^2+Q^2(\df\bar{\theta}^2+\sin^2\bar{\theta}\df\phi^2)\ ,\\
A_\phi&=Qe^{\gamma/2}(1-\cos\bar{\theta})\ ,\label{eq:Aphi}
\end{align}
with 
\begin{align}
Q=\dfrac{Mr_+}{\sqrt{M^2+a^2e^{-\gamma}}}\ .
\end{align}
The charge calculated from (\ref{eq:Komar}) is
\begin{align}
P=e^{\gamma/2}Q\ .
\end{align}
In the limit of large $a$, we have $Q\to M$, the gauge potential is given by (\ref{eq:Aphi}) and the metric goes to
\begin{align}
ds^2=-\left(1+\dfrac{Q}{\rho}\right)^{-2}+\left(1+\dfrac{Q}{\rho}\right)^{2}[\df\rho^2+\rho^2(\df\bar{\theta}^2+\sin^2\bar{\theta}\df\phi^2)]\ ,
\end{align}
which is precisely extremal magnetically charged ModMax BH. For the magnetic field value given by (\ref{eq:mag}), the BH horizon is spherical.  In all other cases,  the presence of another BH together with a conical singularity, cause deviations from spherical symmetry \cite{EMP00}. 

Duality invariance allows for the derivation of the electric version of the solution, with 
\begin{align}
\textbf{A}=-e^{-\gamma}\cos\theta\left(-Dr+3DM-e^{-\gamma}aD^2M-\dfrac{1}{2}e^{-\gamma}aD^2M\sin^2\theta+\dfrac{2aM}{\Sigma}\left(1-\dfrac{1}{2}aDe^{-\gamma}\sin^2\theta  \right)^2   \right) \df t\ ,
\end{align}
while the metric in (\ref{eq:MMdihole}) remains the same, up to the field parameter redefinition $B\to-D$. 
\subsection{Dihole-dilaton-ModMax}
We may extend the result by introducing real scalar fields that are nonminimally coupled to the NLE sector.  An example is the dilaton field, with exponential coupling determined by the parameter $\alpha$. For $\alpha=1$, dilaton fields naturally arise in the low-energy effective action of string theory, while $\alpha=\sqrt{3}$ corresponds to the Kaluza-Klein scenario. The Bonnor-dilaton solution in Maxwell’s electrodynamics for an arbitrary coupling $\alpha$ was first presented in \cite{DAVIDSON1994304}. However, the correct physical interpretation of this solution, along with its embedding into the Melvin universe as a method of regularising conical singularities, was provided in \cite{EMP00}. In this section, we will derive its counterpart within ModMax theory.

The Lagrangian density for NLE-dilaton system is given by 
\begin{align}\label{eq:ldil}
\LL^{(em, dil)}=e^{-2\alpha\Phi}\LL^{(MM)}-\dfrac{1}{2}\nabla_a\Phi\nabla^a\Phi \ ,
\end{align}
where $\Phi$ denotes the dilaton field. When coupled to Einstein’s gravity, (\ref{eq:ldil}) yields the following system of equations for magnetic configurations \cite{Bokulic2023},
\begin{equation}
\left.\begin{aligned}
&R_{ab}=2\nabla_a\Phi\nabla_b\Phi+8\pi e^{-2\alpha\Phi}e^{-\gamma}T_{ab}^{(Max)}\ ,\\
&\Box\Phi+\dfrac{\alpha}{2}e^{-2\alpha\Phi}e^{-\gamma}\FF=0\ ,\\
&\df(e^{-2\alpha\Phi}{\hdg \textbf{F})}=0,\ \df\textbf{F}=0\ .
\end{aligned}\right.
\end{equation}
The dilatonic generalisation of the solution (\ref{eq:MMdihole}) can be derived using an approach similar to the one outlined in the previous section. Bulilding upon a procedure presented in \cite{DAVIDSON1994304, Liang:2001sp}, one can derive the Bonnor-ModMax-dilaton solution,
\begin{equation}
\left.\begin{aligned}\label{eq:BMMD}
\df s^2&=-\left(1-\dfrac{2Mr}{\Sigma}\right)^{\frac{2}{1+\alpha^2}}\df t^2+\dfrac{(1-\frac{2Mr}{\Sigma})^{\frac{2}{1+\alpha^2}}\Sigma^{\frac{4}{1+\alpha^2}}}{(\Delta+(M^2+a^2e^{-\gamma})\text{sin}^2\theta)^{\frac{3-\alpha^2}{1+\alpha^2}}}\left(\dfrac{\df r^2}{\Delta}+\df \theta^2\right)+
\dfrac{\Delta\text{sin}^2\theta}{(1-\frac{2Mr}{\Sigma})^{\frac{2}{1+\alpha^2}}}\df \phi^2\ ,\\\textbf{A}&=\dfrac{2aMr\text{sin}^2\theta}{\sqrt{1+\alpha^2}(\Delta+a^2e^{-\gamma}\text{sin}^2\theta)}\df \phi\ ,\ \ \
\Phi=-\dfrac{\alpha}{1+\alpha^2}\text{Log}\left(1-\dfrac{2Mr}{\Sigma}\right)
\end{aligned}\right.
\end{equation}
By applying a modified Harrison-dilaton transformation, 
\begin{equation}
\left.\begin{aligned}
&g'_{ii}=\lambda^{2/(1+\alpha^2)} g_{ii}\ \  \text{if}\ \ i\neq\phi\ , \  g'_{\phi\phi}=\lambda^{-2/(1+\alpha^2)}g_{\phi\phi}\ ,\ \ \\
A'_\phi=&\dfrac{-2e^{\gamma}(-1+\lambda)-BA_\phi(1+\alpha^2)}{B\lambda(1+\alpha^2)}\ ,\ \  e^{-2\alpha\Phi'}=e^{-2\alpha\Phi}\lambda^{2\alpha^2/(1+\alpha^2)}\ ,\\
\lambda&=\left(1-\dfrac{1}{2}e^{-\gamma}BA_\phi(1+\alpha^2)\right)^2+\dfrac{1}{4}B^2(1+\alpha^2)e^{2\alpha\Phi}e^{-\gamma}f(\rho, z)\ ,
\end{aligned}\right.
\end{equation}
the solution (\ref{eq:BMMD}) can be embedded into the Melvin-ModMax-dilaton universe. The resulting spacetime is then given by
\begin{equation}
\left.\begin{aligned}
\df s^2&=-\Lambda^{\frac{2}{1+\alpha^2}} \df t^2+\dfrac{\Lambda^{\frac{2}{1+\alpha^2}}\Sigma^{\frac{4}{1+\alpha^2}}}{(\Delta+(M^2+a^2e^{-\gamma})\text{sin}^2\theta)^{\frac{3-\alpha^2}{1+\alpha^2}}}\left(\dfrac{\df r^2}{\Delta}+\df \theta^2\right)+\dfrac{\Delta\text{sin}^2\theta}{\Lambda^{\frac{2}{1+\alpha^2}}}\df\phi^2\ , \\
\textbf{A}&=-\frac{-2Mra/\sqrt{1+\alpha^2}+\frac{1}{2}B((r^2-a^2e^{-\gamma})^2+\Delta a^2e^{-\gamma}\text{sin}^2\theta)}{\Lambda\Sigma}\text{sin}^2\theta\df \phi\ ,\ \Phi=-\text{Log}(\Lambda^{\frac{\alpha}{1+\alpha^2}})
\end{aligned}\right.
\end{equation}
where
\begin{align}
\Lambda=\dfrac{\Delta+a^2e^{-\gamma}\text{sin}^2\theta-2\sqrt{1+\alpha^2}BMrae^{-\gamma}\text{sin}^2\theta+(1+\alpha^2)\frac{1}{4}B^2e^{-\gamma}\text{sin}^2\theta((r^2-a^2e^{-\gamma})^2+\Delta a^2e^{-\gamma}\text{sin}^2\theta)}{\Sigma}\ .
\end{align}
The conical singularities are absent for a particular value of the magnetic field,
\begin{align}
B=\dfrac{a\pm\sqrt{a^2+e^{\gamma}M^2}}{\sqrt{1+\alpha^2}Mr_+}\ .
\end{align}
In the limit of large $a$, the solution asymptotes to
\begin{equation}
\left.\begin{aligned}
\df s^2&=-\left(1+\dfrac{Q}{\rho}\right)^{-\frac{2}{1+\alpha^2}}+\left(1+\dfrac{Q}{\rho}\right)^{\frac{2}{1+\alpha^2}}[\df\rho^2+\rho^2(\df\bar{\theta}^2+\sin^2\bar{\theta}\df\phi^2)]\ ,\\
A_\phi&=\dfrac{e^{\gamma/2 }Q(1-\cos\bar{\theta})}{\sqrt{1+\alpha^2}}\ ,\ \ e^\Phi=\left(1+\dfrac{Q}{\rho}  \right)^{\frac{\alpha}{1+\alpha^2}}\ ,
\end{aligned}\right.
\end{equation}
corresponding to the extremal dilatonic ModMax BHs, which were, to the best of our knowledge, previously not discussed in the literature. The near-horizon limit reveals a physical singularity at the poles, as evident from the behaviour of the Ricci scalar,
\begin{align}
R\sim \rho^{-2\alpha^2/(1+\alpha^2)} \ \ \text{as}\ \  \rho\to0\ .
\end{align}
This comes as no surprise, as singular horizons are a well-known feature of extremal electrically or magnetically charged Maxwell-dilaton black holes \cite{GIBBONS1988741}.
For $M=0$ and $a=0$, the solution reduces to the Melvin-ModMax-dilaton universe,
\begin{equation}
\left.\begin{aligned}
ds^2=\Lambda^{\frac{2}{1+\alpha^2}}&(-\df t^2+\df r^2+r^2\df \theta^2)+\Lambda^{-\frac{2}{1+\alpha^2}}r^2\sin^2\theta\df \phi^2,\\
&e^{-2\alpha\Phi}=\Lambda^{\frac{2\alpha^2}{1+\alpha^2}}\, \ A_\phi=-\dfrac{Br^2\sin^2\theta}{2\Lambda}\ ,
\end{aligned}\right.
\end{equation}
where $\Lambda=1+\frac{1}{4}(1+\alpha^2)B^2e^{-\gamma}r^2\sin^2\theta$.
This may be particularly interesting for the cases 
 $\alpha=1$ or $\alpha=\sqrt{3}$ where the model corresponds to Kaluza-Klein gravity or a consistent truncation in heterotic string theory. 
\section{Discussion}\label{sec6}
Owing to the Maxwell-like character of ModMax theory, we were able to partially generalise the Ernst scheme for the EMM system, making its symmetry transformations explicit. The limitation to full generalisation arises from the fact that the similarity with Maxwell’s electrodynamics is preserved only in static spacetimes. This restriction excluded any transformations that would mix real and imaginary parts of gravitational or electromagnetic Ernst potentials. By considering electric and magnetic configurations separately, we have derived two Harrison transformations, which can be used to expand the EMM solution space. As an illustration, this procedure was utilised to derive the novel ModMax-dihole solution, corresponding to two oppositely charged, extremal, magnetically charged ModMax BHs immersed in the magnetic Melvin universe.  By adjusting the value of the magnetic field, the gravitational attraction is balanced and the solution is devoid of conical singularities. A further generalization of the solution, introducing a dilatonic coupling to the ModMax field in the action, was also provided. 
\newline
\indent A puzzling feature arises when applying electromagnetic duality transformation to solutions describing electric (magnetic) BH immersed into electric (magnetic) universe\footnote{This feature is not unique to ModMax theory, but is observed in Maxwell case as well.} (\ref{eq:elel}). The dyonic solution obtained via the duality transformation, which operates on the gauge sector only, will remain static. On the other hand, if we start from an electrically charged BH and place it into a magnetic universe, a nonzero Poynting vector will introduce rotation \cite{Gibbons_2013, PhysRevD.89.044022}.  This can be understood by noting that electromagnetic duality generates a particular subclass of solutions where the electromagnetic fields remain proportional. Consequently,  the Poynting vector vanishes and the resulting metric does not acquire the cross term associated with rotation. An avenue worth exploring would be the construction of a stationary class of solutions with nonvanishing Poynting vector in the ModMax theory. Such solutions would likely exhibit a richer structure, differing significantly from their Maxwell counterparts.
\newline
\indent The presented Ernst mechanism opens the door to generating further new solutions in ModMax electrodynamics. For example, it is expected that the accelerated type I BHs, similar to those in \cite{PhysRevD.108.124025}, could be derived by applying the Harrison transformation to the uncharged C-metric written in the electric LWP ansatz. The conical singularities of C-metric ModMax BHs \cite{BARRIENTOS2022137447} could be regularised with the external electric or magnetic field \cite{10.1063/1.522935, PhysRevD.82.024006}. Relying on the intuition from the Maxwell case, the Ernst formalism could be straightforwardly extended to include particular theories with scalar fields, alongside ModMax electromagnetic fields \cite{PhysRevD.87.084029, PhysRevD.88.104027, PhysRevD.108.024059}. Another possibility is to adapt techniques from \cite{Astorino2012,Barrientos:2024pkt} to include the cosmological constant in the formalism.
\newline
\indent The Ehlers transformation,  another nontrivial symmetry of the Einstein-Maxwell system, necessarily generates a nonstatic spacetime, where rotation is associated with NUT charge or swirling effects. Finding its equivalent in ModMax theory is challenging, since the proportionality between the electromagnetic invariants is generally compromised. Even though the integrability of field equations in Taub-NUT and vortex solutions is preserved \cite{BCPK24}, it remains unclear how to incorporate the Ehlers transformation into the ModMax framework. Thus, the central question remains: is it possible to fully extend the Ernst scheme to ModMax theory, or are there any other NLE theories that allow a simpler description in terms of Ernst potentials? Finding new solutions, with the emphasis on the nonstatic ones, will certainly shed some light on this problem as well.

\begin{acknowledgments}
We thank Adolfo Cisterna and Ivica Smolić for valuable comments and suggestions on the draft of this paper. This work is supported by CIDMA under the FCT Multi-Annual Financing
Program for R\&D Units (UID/04106),
through the Portuguese Foundation for Science and Technology (FCT -- Fundaç\~ao para a Ci\^encia e a Tecnologia), as well as the projects: Horizon Europe staff exchange (SE) programme HORIZON-MSCA2021-SE-01 Grant No. NewFunFiCO-101086251;  2022.04560.PTDC (\url{https://doi.org/10.54499/2022.04560.PTDC}) and 2024.05617.CERN (\url{https://doi.org/10.54499/2024.05617.CERN}).  A. B. acknowledges support from the Croatian Science Foundation, Project No. IP-2020- 02-9614.
\end{acknowledgments}

\begin{appendices}
\section{"Dyonic" Harrison transformation}\label{app}
Since ModMax theory admits static solutions with both electric and magnetic fields\footnote{Several such solutions were reviewed in Section \ref{secMH}, where they were derived by applying either the electric or magnetic Harrison transformation followed by the electromagnetic duality transformation.}, it is worth to explore how to derive them from a single Harrison transformation. This can be achieved by considering a more general setting in which the Harrison parameter $\alpha$ is not strictly real or imaginary,  but complex, with both real and imaginary components.  Consequently, the resulting electromagnetic Ernst potential also becomes complex, giving rise to electric and magnetic components of the gauge potential. However, there are few limitations to this construction. To ensure that the gravitational Ernst potential remains real after the transformation, which is necessary for static spacetimes, we use vacuum seed solutions. Also, we assume that the electromagnetic invariants are proportional and define their ratio as $\FF/\GG=\beta$. After introducing the parameter $w=\text{cosh}\gamma-\dfrac{\beta}{\sqrt{1+\beta^2}}\text{sinh}\gamma$, the Ernst equations assume the form
\begin{equation}
\left.\begin{aligned}\label{eq:ErnstDyonic}
(\Epsilon+w\Phi^*\Phi)\nabla^2\Epsilon=\nabla\Epsilon(\nabla\Epsilon+2w\text{Re}(\Phi^*\nabla\Phi))\ ,\\
(\Epsilon+w\Phi^*\Phi)\nabla^2\Phi=\nabla\Phi(\nabla\Epsilon+2w\text{Re}(\Phi^*\nabla\Phi))\ 
\end{aligned}\right.
\end{equation}
The relations between the Ernst potentials and functions in the electric LWP ansatz are given by
\begin{align}
f=\Epsilon+w\Phi^*\Phi, \hspace{5mm}\Phi=A_t+i\tilde{A}_\phi \ ,\\
\hat{\phi}\times\nabla\tilde{A}_\phi=\dfrac{f}{\rho}\nabla A_\phi ,
\end{align}
while for the magnetic LWP ansatz we have
\begin{align}
f=-\Epsilon-w\Phi^*\Phi, \hspace{5mm}\Phi=\tilde{A}_t-iA_\phi \ ,\\
\hat{\phi}\times\nabla\tilde{A}_t=-\dfrac{f}{\rho}\nabla A_t\ .
\end{align}
The transformation from a vacuum seed solution to the configuration described by the system of equations (\ref{eq:ErnstDyonic}) is
\begin{align}
\Epsilon'=\dfrac{\Epsilon}{1 -\alpha^*\alpha w\Epsilon }, \ \ \Phi'=\dfrac{\alpha\Epsilon }{1-\alpha^*\alpha w\Epsilon }
\end{align}
Dyonic solutions from Section \ref{secMH} can now be derived in a more straightforward manner, eliminating the need to use the electromagnetic duality transformation. For instance,  the dyonic ModMax BH is obtained from the Schwarzschild metric in the electric LWP ansatz after performing similar coordinate transformations as in (\ref{eq:coord1}) and setting $\text{Re}\alpha=Qe^{-\gamma}/(2M)$ and $\text{Im}\alpha=P/(2M)$. By using the Schwarzschild metric expressed in the magnetic LWP ansatz as a seed,  we recover the dyonic ModMax-Melvin-Schwarzschild spactime for $\text{Re}\alpha=e^{-\gamma}D/2$ and $\text{Im}\alpha=-B/2$.
\end{appendices}
\bibliography{refsH}
\end{document}